\documentclass[journal=jacsat,manuscript=article]{achemso}

\usepackage[version=3]{mhchem} 

\author{Hrishit Banerjee}
\affiliation{Yusuf Hamied Department of Chemistry, University of Cambridge, Lensfield Road, CB2 1EW, Cambridge, Cambridgeshire, UK.}
\altaffiliation{School of Science and Engineering, University of Dundee, Nethergate, DD1 4HN, Scotland, UK.}
\email{hb595@cam.ac.uk}
\author{Md. Khaja Nazeeruddin}
\affiliation{Group for Molecular Engineering of Functional Materials, Institute of Chemical Sciences and Engineering, École Polytechnique Fédérale de Lausanne, Valais Wallis, Sion, Switzerland.}
\author{Sudip Chakraborty}
\affiliation{Materials Theory for Energy Scavenging (MATES) Lab, Department of Physics, Harish-Chandra Research Institute(HRI), A CI of Homi Bhabha National Institute (HBNI), Chhatnag Road, Jhunsi, Prayagraj 211019, India}
\email{sudiphys@gmail.com}

\title{Tuning Electronic and Optical Properties of 2D/3D Construction based on Hybrid Perovskites through Interfacial Charge Transfer: Towards Higher Efficiency Solar Cells}

\abbreviations{IR,NMR,UV}

\begin{document}

\begin{tocentry}
    
\end{tocentry}

\begin{abstract}

The 2D/3D construction of hybrid perovskite interfaces is gaining increasing attention due to their enhanced stability towards degradation without compromising the corresponding solar cell efficiency. Much of it is due to the interfacial charge transfer and its consequences on the electronic and optical response of the composite system, which are instrumental in the context of stability and efficiency. In this work, we have considered a case study of an experimentally motivated 2D/3D interface constructed based on Ruddlesden-Popper phases of (A43)$_2$PbI$_4$ and (A43)$_2$MAPb$_2$I$_7$ hybrid perovskites to envisage the unique tuning of electronic and optical properties through the associated charge transfer. The corresponding tuning of the band gap is seen to be related to a unique charge transfer process between the 2D and 3D counterparts of the interface mediated from valence to conduction band edges of the composite. We have found that the optical absorption spectra can also be tuned by the construction of such a hetero-interface and the emergence of a unique two-peak step feature on the absorption edge, which is not present in either the 2D or 3D hybrid perovskites. Formation of the composite is found to increase the spectroscopic limited maximum efficiency for the use of these materials as solar cells from $\approx$ 24\% for individual components to $\approx$ 32\% for the composite hetero-structure.
\end{abstract}
\maketitle 


Perovskite solar cells (PSC) have rapidly emerged as a promising alternative to traditional silicon-based solar cells, due to their high efficiency and low cost\cite{Manser2016, Gangadharan2019, Yang2017, Tsai2016, Bakr2017, Razza2016}.
Hybrid perovskite solar cells have been of immense interest both in experimental \cite{Kojima2009, Snaith2016, Yin2014} as well as theoretical condensed matter \cite{BANERJEE2022183}  and materials research in the energy sector due to their extraordinary performances and ease of fabrication \cite{Gratzel2017, Green2017, Bush2017}. However, the poor device stability due to degradation upon water exposure still impedes its widespread commercialisation \cite{Guangda2015, WANG20162, Bakr2017}. 
The effect of moisture on PSC has received substantial attention on
account of the presence of water under realistic operating conditions of solar cells \cite{Slavney2017, Smith2014}. In the presence of moisture, hydrolysis of the perovskite happens, triggered by the
hygroscopic nature of the material.  Prolonged
exposure to water and high temperature induces the
deterioration of the solar cell electrodes due to reaction with
the byproducts of the perovskite decomposition.  This phenomenon results in a drop in the
photovoltaic performances upon a few hours of operation \cite{Cho2018}.

Among various perovskite materials, 2D/3D architecture of hybrid perovskites has gained increasing attention in recent years for their enhanced stability and performance \cite{Grancini2017, Cho2018, Karabag2023, Shirzadi2023, Lin2018, Tan2017}. These hybrid perovskites consist of both a 3D perovskite structure and a 2D perovskite layer, which forms a protective barrier on top of the 3D structure, leading to improved stability and reduced degradation over time \cite{Grancini2017, yongyoon2018, Zhang2018, Lin2018}. Additionally, the 2D/3D hybrid perovskites offer a unique bandgap tuning capability, allowing for the absorption of a broader range of solar light and thus boosting the overall efficiency of the solar cell \cite{yongyoon2018}. 2D or layered perovskites when combined with 3D perovskites in a 2D/3D hybrid, a synergistic action can be designed to boost efficiency and stability \cite{Koh2018} including high thermal stability \cite{Lin2018}. In particular, 2D/3D composites, obtained by blending standard bulky organic cations (as R component) with the precursor of the 3D perovskite, have been recently embodied in solar cells to push device performances and stability. 

In recent work, the concept of engineering 2D/3D composites, aiming to create an LDP water-repellent sheath, containing a saturated highly fluorinated (fluorous) organic cation designed ad hoc on the top of the 3D perovskite bulk was pushed further forward \cite{Cho2018}. They evaluated the effect of the fluorous perovskite by incorporating the cation in two alternative ways: (a) by direct blending of the LDP- and 3D perovskite precursors and (b) by engineering a controlled in-situ layer-by-layer approach which enables the construction of a clean 3D/2D interface. They incorporated the LDP in two different 3D perovskite compositions, MA$_{0.9}$FA$_{0.1}$PbI$_3$ (MFPI) and Cs$_{0.1}$FA$_{0.74}$MA$_{0.13}$PbI$_{2.48}$Br$_{0.39}$
(CFMPIB). In the first case, the fluorous cation was added
to the MFPI perovskite precursors; in the second case, the
cation was deposited on top of preformed CFMPIB through a
 layer-by-layer passivation approach.  In both cases, a thin layer of fluorous LDP self-assembles on the top surface of the 3D
bulk, forming a water-proof sheath. They envisaged the use of properly designed fluorous ammonium cations to modulate the dimensionality of perovskite materials and template the formation of LDP structures. Indeed, because of their shape and large size, much beyond one of standard MA and FA cations which fit in the voids of the 3D perovskite structure, these cations might act as effective spacers between PbX$_6$ octahedra layers. The robustness of fluorous LDP compared to standard LDP might be enhanced due to the
hydrophobic and solvophobic character of the perfluoroalkyl
residues. Based on these premises, they synthesized the
fluorous cation (CF$_3$)$_3$CO(CH$_2$)$_3$NH$_3^+$ in the form of its iodide salt (named A43 from hereon). Two defined structures, arranged into Ruddlesden-Popper phase of (A43)$_2$PbI$_4$ (n = 1) and (A43)$_2$MAPb$_2$I$_7$ (n =2), as derived by X-ray diffraction (XRD) measurement, were obtained. In the case of (A43)$_2$PbI$_4$ perovskite structure, bilayers of bulky A43 cations of 20.65 $\AA$ in length intercalate in between monolayers of PbI$_6$ octahedra. For (A43)$_2$MAPb$_2$I$_7$, bilayers of A43 cations intercalate in between bilayers of PbI$_6$ octahedra in which MA cations are confined with the distance between the MA cations in two contiguous inorganic slabs being 26.64 $\AA$.

Motivated by the success of this 2D/3D perovskite interface in overcoming moisture-based degradation in this work we examine the properties of this particularly interesting hybrid perovskite interface. Using first-principles density-functional theory (DFT) calculations we examine the basic electronic structure, the charge transfer and optical absorption properties of this experimentally designed 2D/3D hybrid perovskite interface. We find that there is a definitive charge transfer occurring between the 2D to 3D structures and this leads to a tunable optical band gap and tunable optical spectra. Our study concludes that the construction of a 2D/3D hybrid perovskite heterostructure not only leads to the protection of the composite from moisture-based degradation but leads to extra tunability properties in terms of band gaps and optical absorption as well as an increase in maximum efficiency of solar cells.

We have systematically determined the electronic structure, optical properties and charge density distribution to envisage the possible charge transfer mechanism as obtained
in the experimental investigations for the interface of 2D/3D hybrid Perovskites: (A43)$_2$PbI$_4$ and
(A43)$_2$MAPb$_2$I$_7$. We have investigated the electronic structure of the individual systems i.e.
(A43)$_2$PbI$_4$ and (A43)$_2$MAPb$_2$I$_7$  and the corresponding interface constructed between the 3D and 2D
nanostructures.  We describe the results obtained in this section.

\begin{figure}
    \centering
    \includegraphics[width=\columnwidth]{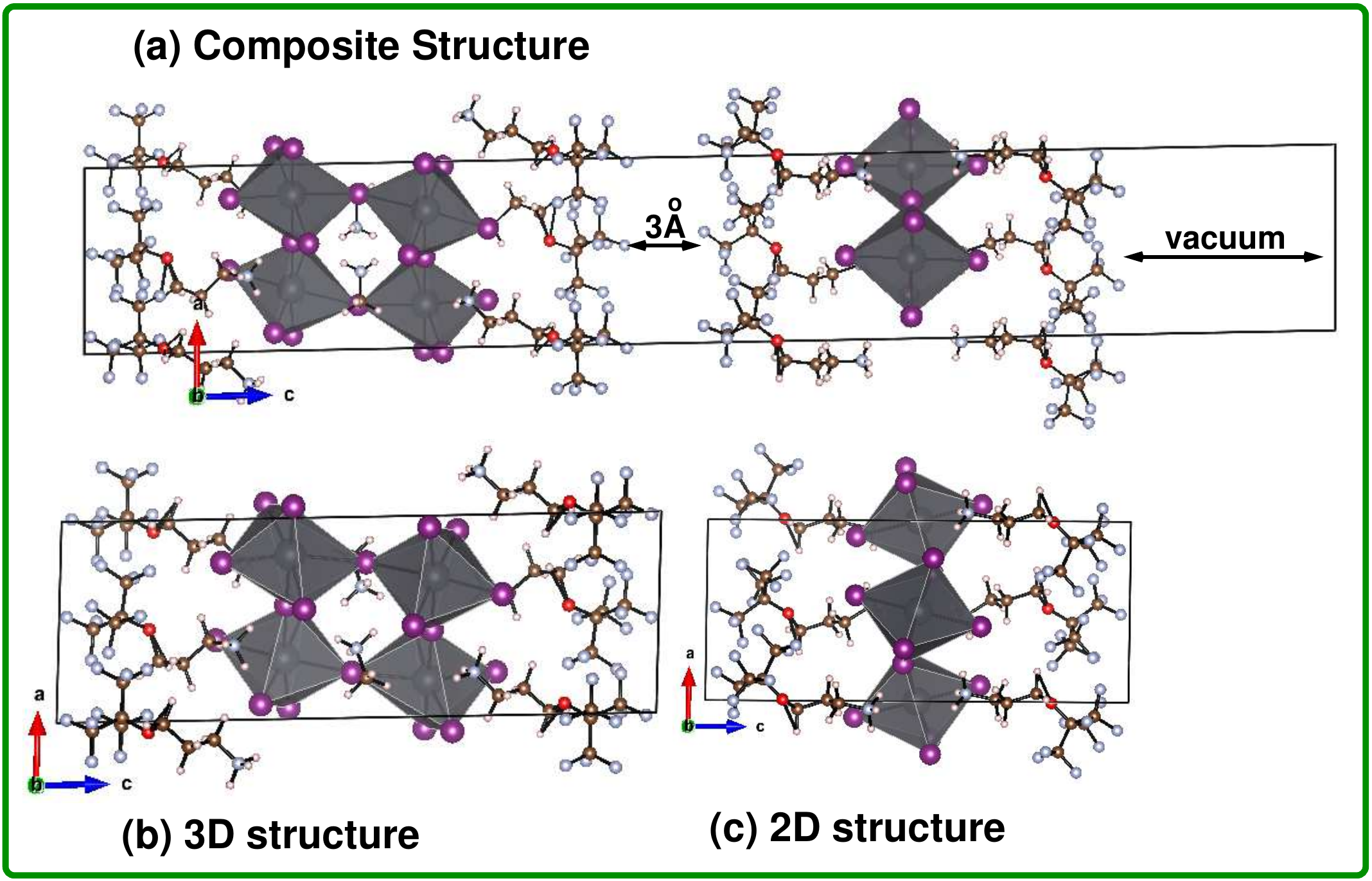}
    \caption{In this figure, we show the structures of the 3D perovskite and the 2D perovskite in the bottom panel and the composite structure with a typical 3eV van der Waals gap in the top panel. The respective PbI$_6$ octahedra are marked in grey.}
    \label{fig1}
\end{figure}

In Figure 1 we show first the crystal structure of the two hybrid perovskites and their composite
structure, where the minimum energy and minimum force configurations of all three systems have been obtained from rigorous first-principles electronic structure calculations. As in the experimental studies the 3D structure is considered as a substrate, by fixing the coordinates of the base perovskite layer in geometry optimisation, mimicking a substrate effect, on which the 2D structure is grown at a
separation of roughly 3$\AA$ between the two structures. This separation corresponds to a typical van der
Waals like distance, and this separation is the optimum distance, at which the two structures are stabilised. This has been confirmed after consideration of several other possible
distances of separation between the two structures, and this was found to be the lowest energy composite structure. We have considered a vacuum of 20 $\AA$ on top of the 2D structure to nullify the interaction of the periodic images of the surface with the substrate along the Z direction.

\begin{figure}
    \centering
    \includegraphics[width=\columnwidth]{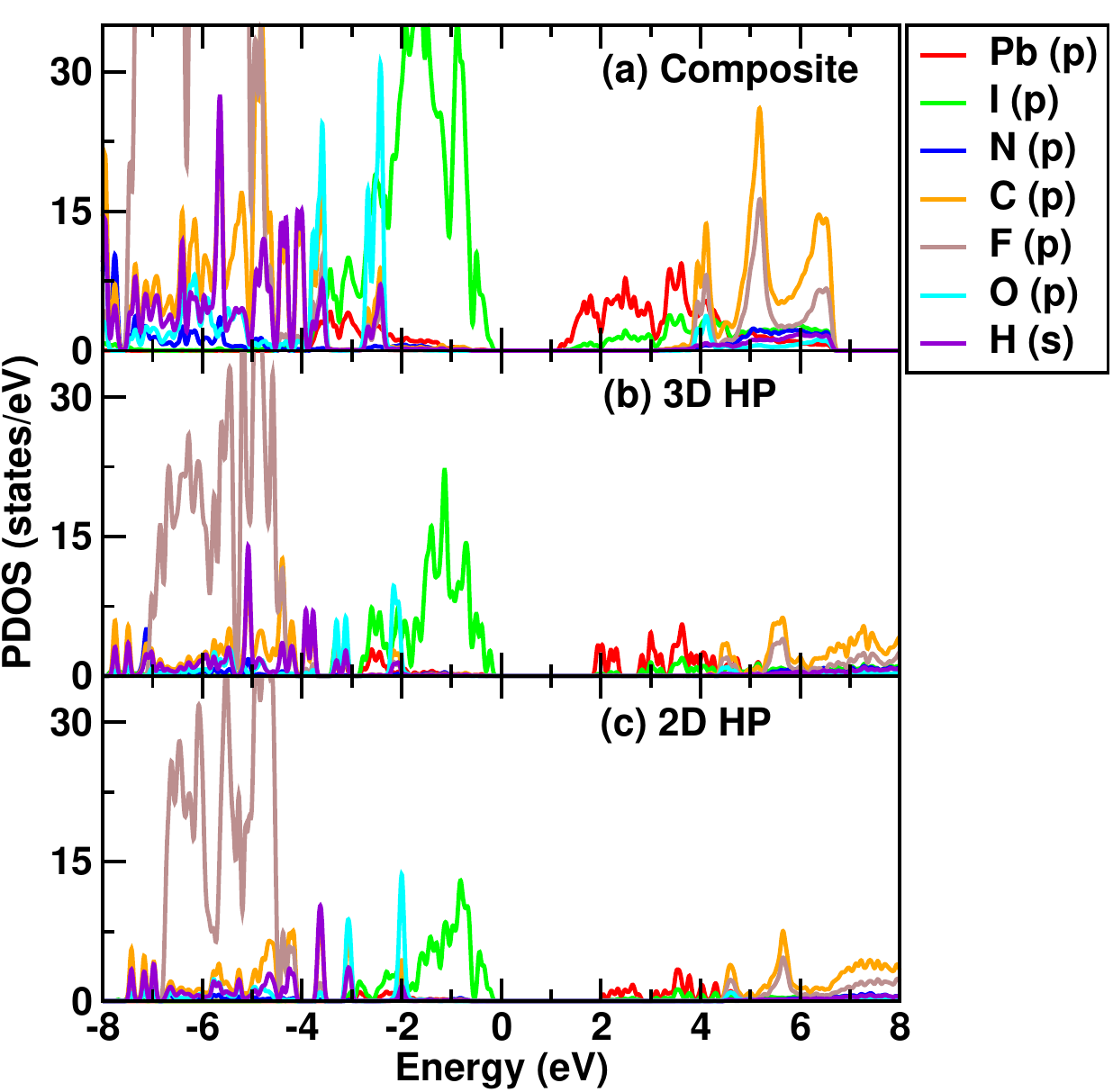}
    \caption{In this figure we show the partial DOS of the composite, the 3D and the 2D structures respectively.}
    \label{fig2}
\end{figure}

In Figure 2, the projected density of states of the individual structures and the composite
structure have been depicted, which primarily explains the electronic properties in the
individual and composite structures. We have shown the projected density of states for elucidation of the contribution of each element and their associated signatures in the valence band
and conduction band regime of the individual and composite structures. In both the 2D and the 3D
system, the dominant contribution close to the Fermi energy arises from the I-p orbitals and with
a small contribution from oxygen and other p-band elements. We have also seen the dominant
contribution of iodine in the composite structure. 
An interesting point to be noted from the projected DOS is that the band gaps in the case of the 3D and the 2D structures are respectively $\sim$1.9eV and $\sim$ 2eV. However, in the composite structure, we obtain a band gap of $\sim$ 1.0eV. This points towards a very illuminating finding that by the construction of the composite structure of a 2D coating on top of the 3D perovskite, the band gap changes and this may be a method of tunability of band gaps, which is an extremely important factor in the construction of perovskite solar cells. We shall discuss this further in the section where we discuss optical absorption spectra where this has greater importance.

We can also relate the projected density of states
with the associated elemental charges as determined by doing a Bader charge analysis. We have also envisaged the charge density distribution of the pristine systems and the
corresponding composite structure, by calculation of partial charge densities as depicted in Figure 3. Comparing the overall charge per
formula unit one finds that whereas the 2D structure has 4.28 e/f.u. and the 3D structure has 8.26
e/f.u., there is an overall charge content of 12.28 e/f.u. found in the composite structure. It is easy to
see that by bringing the structures together there is a transfer of overall charge between the 2D and the
3D structures from the valence to the conduction bands, which also leads to subsequent tunability of optical properties which are dependent on valence to conduction band transitions. 

Subsequently, to illustrate further where the charge transfer from the valence to the conduction band takes place, and which species contributes the most to it, we explore the individual contributions of the dominating ions in the vicinity of the Fermi level, which is eventually iodine (I). We have found that in the 2D case, the
contribution of the I-p charge is 2.74 e/atom, while in the case of the 3D system, the associated charge is 2.77 e/atom. We also need to look into the individual charges in the composite structure from the associated I p orbitals respectively with the 2D and 3D structures. We have found that the iodine atoms from the 3D part of the structure have a valence band contribution of 2.78 e/atom, while that from the 2D counterpart has a contribution of 2.88 e/atom. So this shows two different things. Firstly we observe that I p orbitals are electron-rich in the case of the composite compared to the individual 2D and 3d structures. However, the 2D counterpart is more electron-rich than the 3D counterpart in the composite structure. However, due to the shorter Pb-I bond length of the 2D structure, there is higher hybridisation between the Pb and I conduction orbitals. This eventually leads to a 3D to 2D charge transfer mediated by a valence band to conduction band charge transfer. This idea is reinforced by the partial DOS where the composite structure has more delocalised hybridised Pb-I conduction bands compared to more localised conduction bands in the individual structures.  
Thus there is clear evidence of the charge transfer that has occurred between (A43)$_2$PbI$_4$ and (A43)$_2$MAPb$_2$I$_7$ mediated by a valence to conduction band charge transfer and a corresponding tuning of the band gap. Our theoretical study supports the experimental signatures of charge redistribution at the interface of 3D and 2D structures at the time of the interface
reconstruction.

\begin{figure}
    \centering
    \includegraphics[width=\columnwidth]{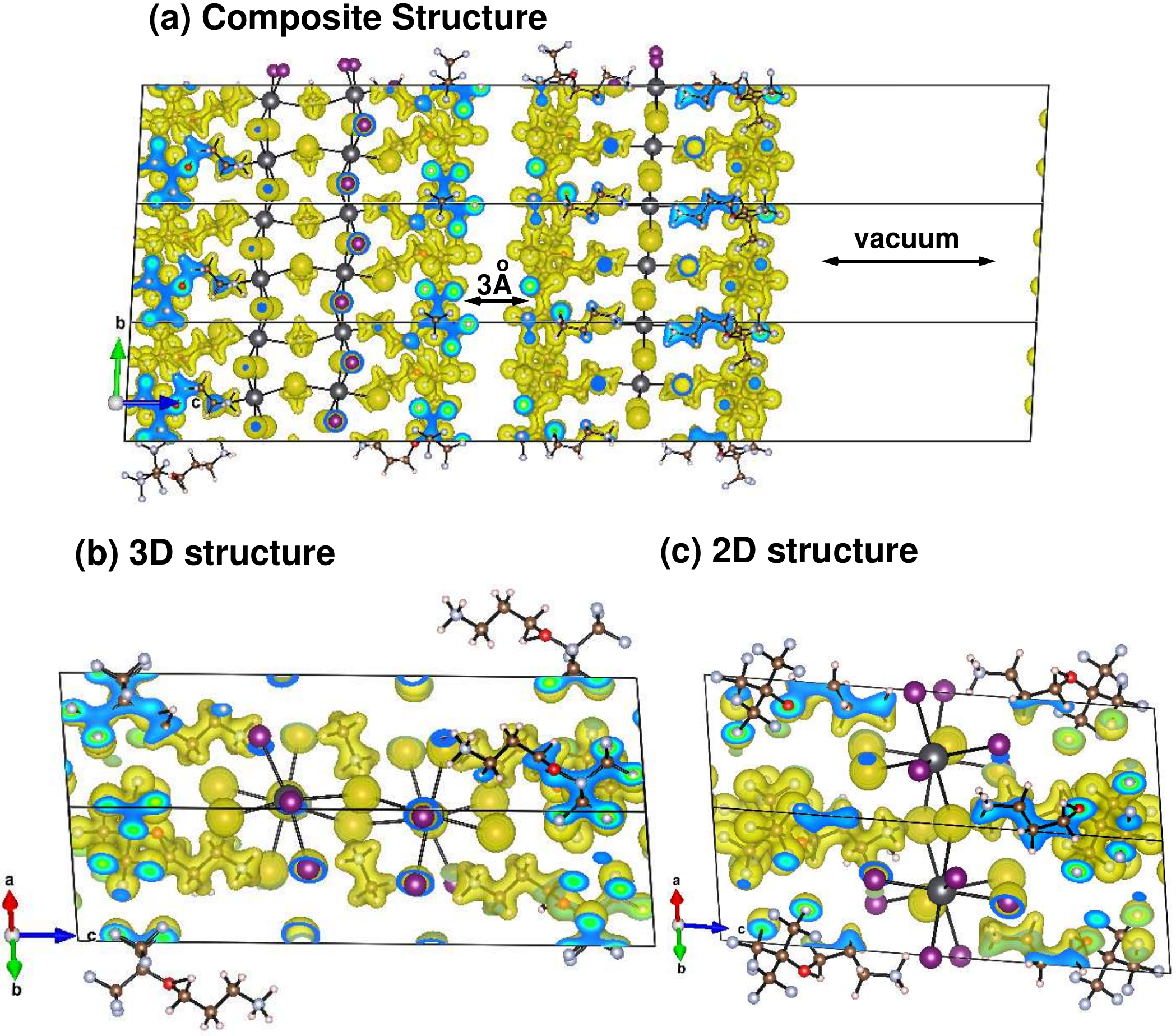}
    \caption{The figure shows the charge densities of the 3D and 2D structures in the bottom panel and the charge density of the composite in the top panel. 
    }
    \label{fig4}
\end{figure}

Finally, we examine the optical absorption spectra, as depicted in Figure 4, of the individual structures and the corresponding composite system, to envisage the effect of heterostructure interface formation on the optical properties of the 2D and 3D hybrid perovskites. This is extremely important as the optical absorption spectra of any system under illumination can be expressed as the variation of absorption cross-section as a function of photon energy, which is one of the prime factors in defining the photo-conversion efficiency of a solar cell\cite{Blancon2017}.
Interestingly, if we observe the corresponding photon energy associated with the start of the absorption cross-section peak, we can see a clear change while going from the individual systems to the construction of the composite system. One can relate this to the projected density of states of the considered systems, as the band gap of the individual system eventually
governs the absorption peak in optical spectra. The change in the band gap as discussed previously corresponds to the electronic structure of the composite system, which originated from the charge transfer between
the 2D and 3D hybrid perovskites mediated through the valence to conduction band charge transfer, therefore the shift in the optical absorption spectra is well connected to the charge transfer mechanism of the interface.
However, there are several more things to be considered here. Firstly the composite structure shows a double peak step feature at the absorption edge between 2.3-2.5eV which is not seen in either of the individual absorption edges. A single peak step feature is seen in the 2D structure around 3 eV but no such peak step feature is seen in the 3D structure on the absorption edge. Thus this feature on the absorption edge which is the most important part of the optical absorption spectra in the context of solar cell-based application of perovskites, is completely new and forms due to the formation of the interface and can be attributed to the valence to conduction band charge transfer phenomena observed. This shows the tunability of the absorption edge which is of prime importance for solar cell applications.
This also points towards the tunability of band gap and absorption cross section by the construction of the 2D/3D interface which is a highly desirable property in solar cell materials. Although the higher energy absorption peak features in the composite mimics primarily the absorption peak features of the 3D structure as expected it is interesting to note that they are modulated by the presence of the 2D coating. It can be observed from Figure 4 that in terms of higher energy absorption features the broad second peak in the 3D bulk structures around 8eV becomes much narrower and sharply defined and now there is a clear 2 peak feature in the absorption spectra of the composite structure however no such feature exists in case of the 2D or the 3D structure individually, again reinforcing the idea of tunability of the absorption spectra by construction of the hetero-interface.

Finally, we have calculated the spectroscopic limited maximum efficiency (SLME) based on the method proposed by Yu et al \cite{Yu2012}. This is calculated directly from the absorbance data which is obtained from the optical absorption data. As shown in Figure 6, We find SLME of 24.1\% and and 24.5\% for the 2D and 3D structures respectively. For the composite structure, we find an SLME of 31.6\% which shows an increase in efficiency from the component 2D and 3D structures.

\begin{figure}
    \centering
    \includegraphics[width=\columnwidth]{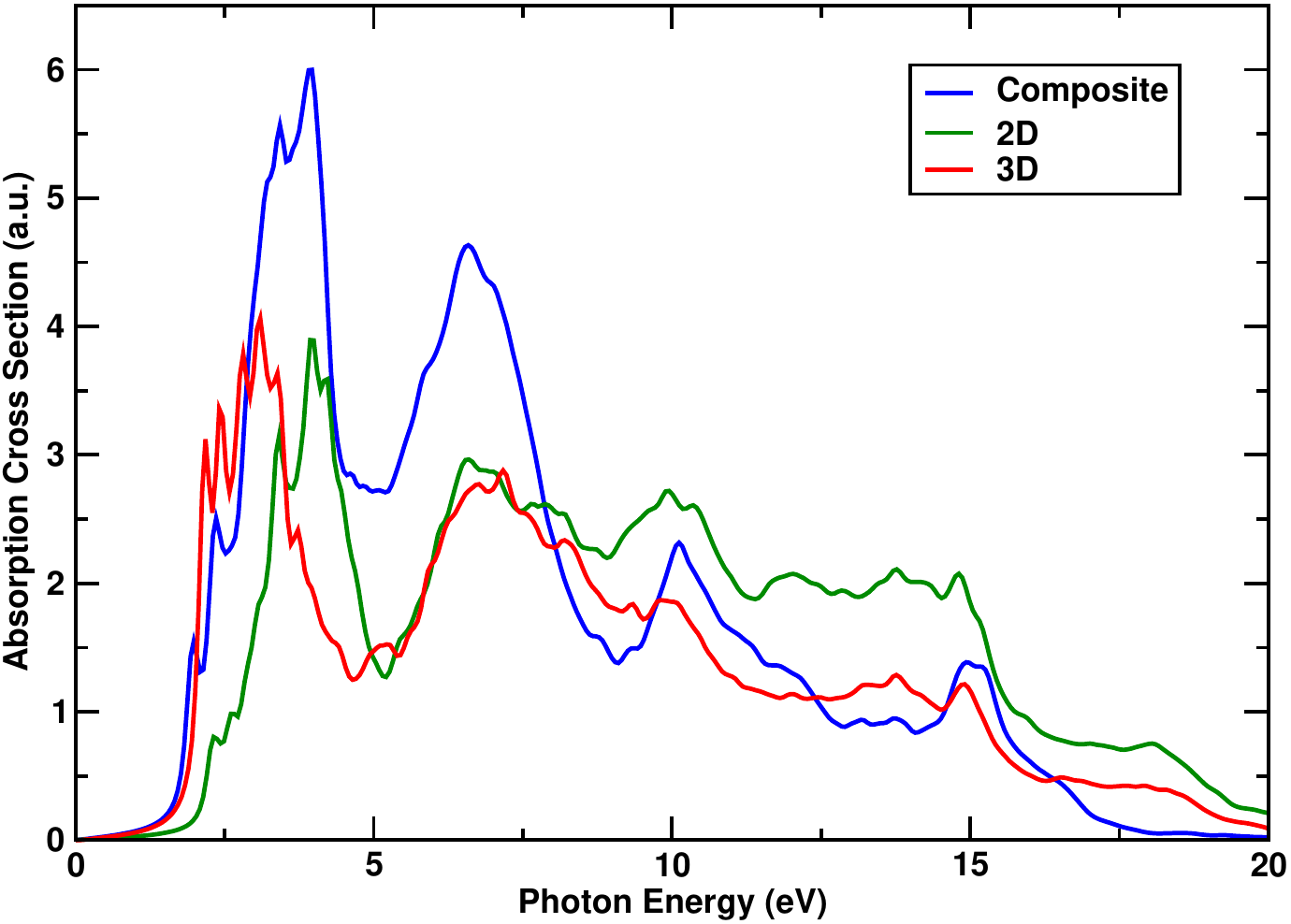}
    \caption{Figure showing optical absorption spectra at the level of independent particle approximation. A unique double peak feature is seen in the optical absorption of the composite material which is not seen in either of the bulk materials.}
    \label{fig5}
\end{figure}

\begin{figure}
    \centering
    \includegraphics[width=\columnwidth]{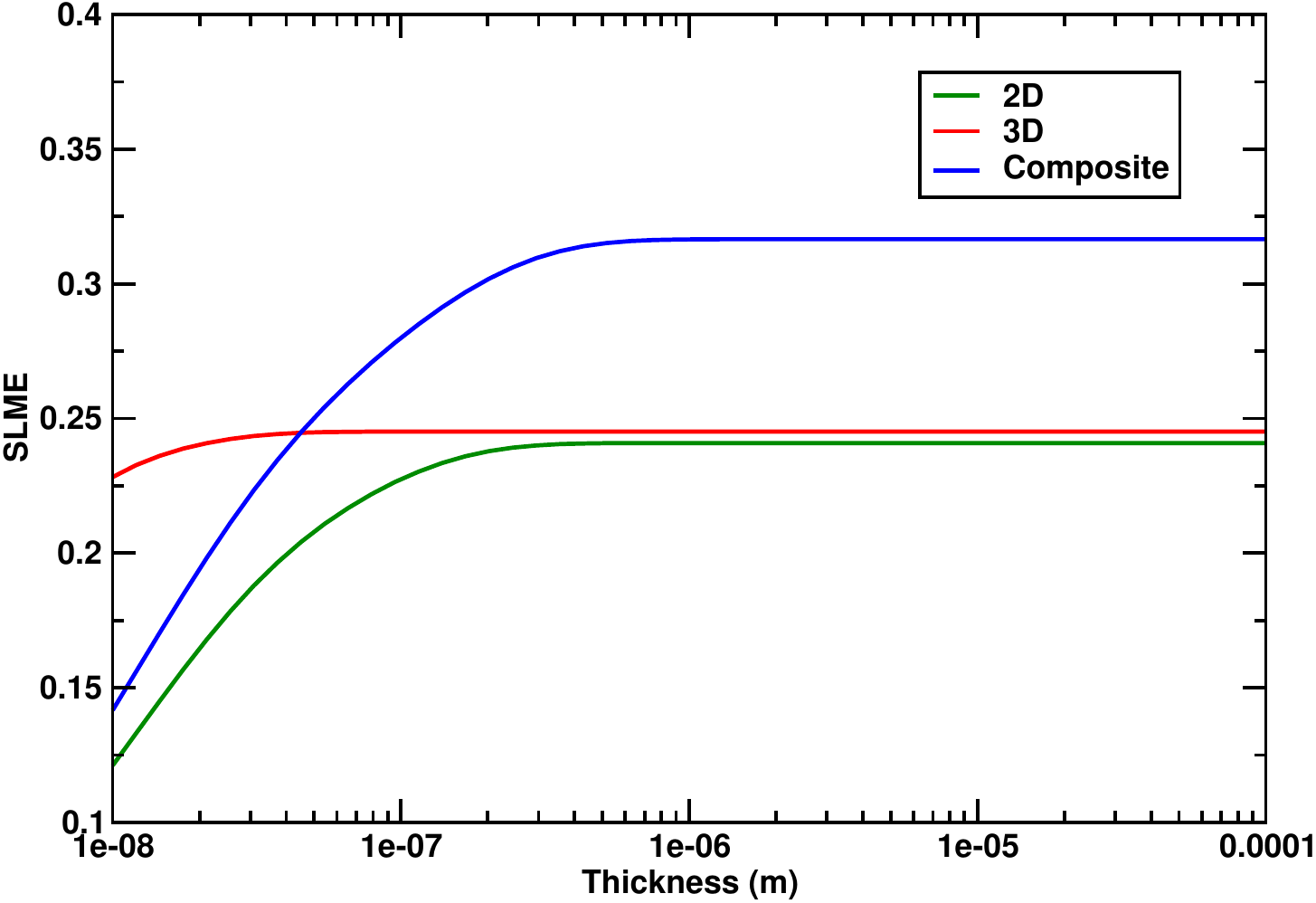}
    \caption{Figure showing Spectroscopic Limited Maximum Efficiency as a function of thickness obtained by the method prescribed by Yu et al \cite{Yu2012}. }
    \label{fig6}
\end{figure}

In conclusion in this work, we have envisaged the tuning of electronic structure and optical properties of 2D/3D construction of RP phases of hybrid perovskite through the interfacial charge transfer. Considering the 3D structure as the matrix and having a single layer of 2D hybrid perovskite constructed on top of the 3D structure, we have found the most stable hetero-structure forms at a van der Waals distance of 3$\AA$ between the 3D and the 2D structures. From our electronic structure analysis, we have found that the majority of contribution near Fermi energy comes from I-5p orbitals hybridised with Pb-6p orbitals. We have also found that even though the 2D and 3D structures individually have a band gap of around 1.9-2eV, the composite structure has a band gap of 1.0eV, which indicates that the band gap was modified by the formation of the hetero-structure. We have performed a Bader charge analysis to explore the corresponding charge transfer mechanism at play which eventually modifies the band gap. The charge transfer occurs from I-5p orbitals in the 3D structure to the 2D structure mediated by an inherent charge transfer from the valence to the conduction band and an eventual increase in Pb-6p - I-5p hybridisation. We have found that the charge transfer mechanism has a profound impact on the optical absorption spectra, where a tunable absorption edge is seen with a novel two-peak step feature along the absorption edge which is not seen in either of the 2D or 3D hybrid perovskites. This is particularly important in terms of tunability of optical absorption properties and applications of these 2D/3D hybrid perovskite hetero-interfaces as solar cells. We also find an increase in the efficiency of the solar cell in the composite system compared to the individual 2D and 3D systems. Our current investigation thus sheds
light on the charge transfer phenomena between the 2D and 3D hybrid perovskite systems when constructed in the interface form, which not only stabilises the hetero-structure but also governs the electronic structures in terms of projected density of states and the tunable optical properties in
terms of absorption cross-section. Our study is expected to give rise to further experimental studies examining the tunable optical absorption spectra in 2D/3D hybrid perovskite interfaces.

\bibliography{achemso}

\providecommand{\latin}[1]{#1}
\makeatletter
\providecommand{\doi}
  {\begingroup\let\do\@makeother\dospecials
  \catcode`\{=1 \catcode`\}=2 \doi@aux}
\providecommand{\doi@aux}[1]{\endgroup\texttt{#1}}
\makeatother
\providecommand*\mcitethebibliography{\thebibliography}
\csname @ifundefined\endcsname{endmcitethebibliography}
  {\let\endmcitethebibliography\endthebibliography}{}
\begin{mcitethebibliography}{29}
\providecommand*\natexlab[1]{#1}
\providecommand*\mciteSetBstSublistMode[1]{}
\providecommand*\mciteSetBstMaxWidthForm[2]{}
\providecommand*\mciteBstWouldAddEndPuncttrue
  {\def\EndOfBibitem{\unskip.}}
\providecommand*\mciteBstWouldAddEndPunctfalse
  {\let\EndOfBibitem\relax}
\providecommand*\mciteSetBstMidEndSepPunct[3]{}
\providecommand*\mciteSetBstSublistLabelBeginEnd[3]{}
\providecommand*\EndOfBibitem{}
\mciteSetBstSublistMode{f}
\mciteSetBstMaxWidthForm{subitem}{(\alph{mcitesubitemcount})}
\mciteSetBstSublistLabelBeginEnd
  {\mcitemaxwidthsubitemform\space}
  {\relax}
  {\relax}

\bibitem[Manser \latin{et~al.}(2016)Manser, Christians, and Kamat]{Manser2016}
Manser,~J.~S.; Christians,~J.~A.; Kamat,~P.~V. Intriguing Optoelectronic
  Properties of Metal Halide Perovskites. \emph{Chemical Reviews}
  \textbf{2016}, \emph{116}, 12956--13008\relax
\mciteBstWouldAddEndPuncttrue
\mciteSetBstMidEndSepPunct{\mcitedefaultmidpunct}
{\mcitedefaultendpunct}{\mcitedefaultseppunct}\relax
\EndOfBibitem
\bibitem[Thrithamarassery~Gangadharan and Ma(2019)Thrithamarassery~Gangadharan,
  and Ma]{Gangadharan2019}
Thrithamarassery~Gangadharan,~D.; Ma,~D. Searching for stability at lower
  dimensions: current trends and future prospects of layered perovskite solar
  cells. \emph{Energy Environ. Sci.} \textbf{2019}, \emph{12}, 2860--2889\relax
\mciteBstWouldAddEndPuncttrue
\mciteSetBstMidEndSepPunct{\mcitedefaultmidpunct}
{\mcitedefaultendpunct}{\mcitedefaultseppunct}\relax
\EndOfBibitem
\bibitem[Yang \latin{et~al.}(2017)Yang, Park, Jung, Jeon, Kim, Lee, Shin, Seo,
  Kim, Noh, and Seok]{Yang2017}
Yang,~W.~S.; Park,~B.-W.; Jung,~E.~H.; Jeon,~N.~J.; Kim,~Y.~C.; Lee,~D.~U.;
  Shin,~S.~S.; Seo,~J.; Kim,~E.~K.; Noh,~J.~H.; Seok,~S.~I. Iodide management
  in formamidinium-lead-halide–based perovskite layers for efficient solar
  cells. \emph{Science} \textbf{2017}, \emph{356}, 1376--1379\relax
\mciteBstWouldAddEndPuncttrue
\mciteSetBstMidEndSepPunct{\mcitedefaultmidpunct}
{\mcitedefaultendpunct}{\mcitedefaultseppunct}\relax
\EndOfBibitem
\bibitem[Tsai \latin{et~al.}(2016)Tsai, Nie, Blancon, Stoumpos, Asadpour,
  Harutyunyan, Neukirch, Verduzco, Crochet, Tretiak, Pedesseau, Even, Alam,
  Gupta, Lou, Ajayan, Bedzyk, Kanatzidis, and Mohite]{Tsai2016}
Tsai,~H. \latin{et~al.}  High-efficiency two-dimensional Ruddlesden--Popper
  perovskite solar cells. \emph{Nature} \textbf{2016}, \emph{536},
  312--316\relax
\mciteBstWouldAddEndPuncttrue
\mciteSetBstMidEndSepPunct{\mcitedefaultmidpunct}
{\mcitedefaultendpunct}{\mcitedefaultseppunct}\relax
\EndOfBibitem
\bibitem[Bakr and Mohammed(2017)Bakr, and Mohammed]{Bakr2017}
Bakr,~O.~M.; Mohammed,~O.~F. Powering up perovskite photoresponse.
  \emph{Science} \textbf{2017}, \emph{355}, 1260--1261\relax
\mciteBstWouldAddEndPuncttrue
\mciteSetBstMidEndSepPunct{\mcitedefaultmidpunct}
{\mcitedefaultendpunct}{\mcitedefaultseppunct}\relax
\EndOfBibitem
\bibitem[Razza \latin{et~al.}(2016)Razza, Castro-Hermosa, Di~Carlo, and
  Brown]{Razza2016}
Razza,~S.; Castro-Hermosa,~S.; Di~Carlo,~A.; Brown,~T.~M. {Research Update:
  Large-area deposition, coating, printing, and processing techniques for the
  upscaling of perovskite solar cell technology}. \emph{APL Materials}
  \textbf{2016}, \emph{4}, 091508\relax
\mciteBstWouldAddEndPuncttrue
\mciteSetBstMidEndSepPunct{\mcitedefaultmidpunct}
{\mcitedefaultendpunct}{\mcitedefaultseppunct}\relax
\EndOfBibitem
\bibitem[Kojima \latin{et~al.}(2009)Kojima, Teshima, Shirai, and
  Miyasaka]{Kojima2009}
Kojima,~A.; Teshima,~K.; Shirai,~Y.; Miyasaka,~T. Organometal Halide
  Perovskites as Visible-Light Sensitizers for Photovoltaic Cells.
  \emph{Journal of the American Chemical Society} \textbf{2009}, \emph{131},
  6050--6051\relax
\mciteBstWouldAddEndPuncttrue
\mciteSetBstMidEndSepPunct{\mcitedefaultmidpunct}
{\mcitedefaultendpunct}{\mcitedefaultseppunct}\relax
\EndOfBibitem
\bibitem[McMeekin \latin{et~al.}(2016)McMeekin, Sadoughi, Rehman, Eperon,
  Saliba, Hörantner, Haghighirad, Sakai, Korte, Rech, Johnston, Herz, and
  Snaith]{Snaith2016}
McMeekin,~D.~P.; Sadoughi,~G.; Rehman,~W.; Eperon,~G.~E.; Saliba,~M.;
  Hörantner,~M.~T.; Haghighirad,~A.; Sakai,~N.; Korte,~L.; Rech,~B.;
  Johnston,~M.~B.; Herz,~L.~M.; Snaith,~H.~J. A mixed-cation lead mixed-halide
  perovskite absorber for tandem solar cells. \emph{Science} \textbf{2016},
  \emph{351}, 151--155\relax
\mciteBstWouldAddEndPuncttrue
\mciteSetBstMidEndSepPunct{\mcitedefaultmidpunct}
{\mcitedefaultendpunct}{\mcitedefaultseppunct}\relax
\EndOfBibitem
\bibitem[Yin \latin{et~al.}(2014)Yin, Shi, and Yan]{Yin2014}
Yin,~W.-J.; Shi,~T.; Yan,~Y. Unique Properties of Halide Perovskites as
  Possible Origins of the Superior Solar Cell Performance. \emph{Advanced
  Materials} \textbf{2014}, \emph{26}, 4653--4658\relax
\mciteBstWouldAddEndPuncttrue
\mciteSetBstMidEndSepPunct{\mcitedefaultmidpunct}
{\mcitedefaultendpunct}{\mcitedefaultseppunct}\relax
\EndOfBibitem
\bibitem[Banerjee \latin{et~al.}(2022)Banerjee, Kaur, Nazeeruddin, and
  Chakraborty]{BANERJEE2022183}
Banerjee,~H.; Kaur,~J.; Nazeeruddin,~M.; Chakraborty,~S. Tuning paradigm of
  external stimuli driven electronic, optical and magnetic properties in hybrid
  perovskites and metal organic complexes. \emph{Materials Today}
  \textbf{2022}, \emph{60}, 183--200\relax
\mciteBstWouldAddEndPuncttrue
\mciteSetBstMidEndSepPunct{\mcitedefaultmidpunct}
{\mcitedefaultendpunct}{\mcitedefaultseppunct}\relax
\EndOfBibitem
\bibitem[Gr{\"a}tzel(2017)]{Gratzel2017}
Gr{\"a}tzel,~M. The Rise of Highly Efficient and Stable Perovskite Solar Cells.
  \emph{Accounts of Chemical Research} \textbf{2017}, \emph{50}, 487--491\relax
\mciteBstWouldAddEndPuncttrue
\mciteSetBstMidEndSepPunct{\mcitedefaultmidpunct}
{\mcitedefaultendpunct}{\mcitedefaultseppunct}\relax
\EndOfBibitem
\bibitem[Green and Ho-Baillie(2017)Green, and Ho-Baillie]{Green2017}
Green,~M.~A.; Ho-Baillie,~A. Perovskite Solar Cells: The Birth of a New Era in
  Photovoltaics. \emph{ACS Energy Letters} \textbf{2017}, \emph{2},
  822--830\relax
\mciteBstWouldAddEndPuncttrue
\mciteSetBstMidEndSepPunct{\mcitedefaultmidpunct}
{\mcitedefaultendpunct}{\mcitedefaultseppunct}\relax
\EndOfBibitem
\bibitem[Bush \latin{et~al.}(2017)Bush, Palmstrom, Yu, Boccard, Cheacharoen,
  Mailoa, McMeekin, Hoye, Bailie, Leijtens, Peters, Minichetti, Rolston,
  Prasanna, Sofia, Harwood, Ma, Moghadam, Snaith, Buonassisi, Holman, Bent, and
  McGehee]{Bush2017}
Bush,~K.~A. \latin{et~al.}  23.6{\%}-efficient monolithic perovskite/silicon
  tandem solar cells with improved stability. \emph{Nature Energy}
  \textbf{2017}, \emph{2}, 17009\relax
\mciteBstWouldAddEndPuncttrue
\mciteSetBstMidEndSepPunct{\mcitedefaultmidpunct}
{\mcitedefaultendpunct}{\mcitedefaultseppunct}\relax
\EndOfBibitem
\bibitem[Niu \latin{et~al.}(2015)Niu, Guo, and Wang]{Guangda2015}
Niu,~G.; Guo,~X.; Wang,~L. Review of recent progress in chemical stability of
  perovskite solar cells. \emph{J. Mater. Chem. A} \textbf{2015}, \emph{3},
  8970--8980\relax
\mciteBstWouldAddEndPuncttrue
\mciteSetBstMidEndSepPunct{\mcitedefaultmidpunct}
{\mcitedefaultendpunct}{\mcitedefaultseppunct}\relax
\EndOfBibitem
\bibitem[Wang \latin{et~al.}(2016)Wang, Wright, Elumalai, and Uddin]{WANG20162}
Wang,~D.; Wright,~M.; Elumalai,~N.~K.; Uddin,~A. Stability of perovskite solar
  cells. \emph{Solar Energy Materials and Solar Cells} \textbf{2016},
  \emph{147}, 255--275\relax
\mciteBstWouldAddEndPuncttrue
\mciteSetBstMidEndSepPunct{\mcitedefaultmidpunct}
{\mcitedefaultendpunct}{\mcitedefaultseppunct}\relax
\EndOfBibitem
\bibitem[Slavney \latin{et~al.}(2017)Slavney, Smaha, Smith, Jaffe, Umeyama, and
  Karunadasa]{Slavney2017}
Slavney,~A.~H.; Smaha,~R.~W.; Smith,~I.~C.; Jaffe,~A.; Umeyama,~D.;
  Karunadasa,~H.~I. Chemical Approaches to Addressing the Instability and
  Toxicity of Lead--Halide Perovskite Absorbers. \emph{Inorganic Chemistry}
  \textbf{2017}, \emph{56}, 46--55\relax
\mciteBstWouldAddEndPuncttrue
\mciteSetBstMidEndSepPunct{\mcitedefaultmidpunct}
{\mcitedefaultendpunct}{\mcitedefaultseppunct}\relax
\EndOfBibitem
\bibitem[Smith \latin{et~al.}(2014)Smith, Hoke, Solis-Ibarra, McGehee, and
  Karunadasa]{Smith2014}
Smith,~I.~C.; Hoke,~E.~T.; Solis-Ibarra,~D.; McGehee,~M.~D.; Karunadasa,~H.~I.
  A Layered Hybrid Perovskite Solar-Cell Absorber with Enhanced Moisture
  Stability. \emph{Angewandte Chemie International Edition} \textbf{2014},
  \emph{53}, 11232--11235\relax
\mciteBstWouldAddEndPuncttrue
\mciteSetBstMidEndSepPunct{\mcitedefaultmidpunct}
{\mcitedefaultendpunct}{\mcitedefaultseppunct}\relax
\EndOfBibitem
\bibitem[Cho \latin{et~al.}(2018)Cho, Zhang, Orlandi, Cavazzini, Zimmermann,
  Lesch, Tabet, Pozzi, Grancini, and Nazeeruddin]{Cho2018}
Cho,~K.~T.; Zhang,~Y.; Orlandi,~S.; Cavazzini,~M.; Zimmermann,~I.; Lesch,~A.;
  Tabet,~N.; Pozzi,~G.; Grancini,~G.; Nazeeruddin,~M.~K. Water-Repellent
  Low-Dimensional Fluorous Perovskite as Interfacial Coating for 20\% Efficient
  Solar Cells. \emph{Nano Letters} \textbf{2018}, \emph{18}, 5467--5474, PMID:
  30134112\relax
\mciteBstWouldAddEndPuncttrue
\mciteSetBstMidEndSepPunct{\mcitedefaultmidpunct}
{\mcitedefaultendpunct}{\mcitedefaultseppunct}\relax
\EndOfBibitem
\bibitem[Grancini \latin{et~al.}(2017)Grancini, Rold{\'a}n-Carmona, Zimmermann,
  Mosconi, Lee, Martineau, Narbey, Oswald, De~Angelis, Graetzel, and
  Nazeeruddin]{Grancini2017}
Grancini,~G.; Rold{\'a}n-Carmona,~C.; Zimmermann,~I.; Mosconi,~E.; Lee,~X.;
  Martineau,~D.; Narbey,~S.; Oswald,~F.; De~Angelis,~F.; Graetzel,~M.;
  Nazeeruddin,~M.~K. One-Year stable perovskite solar cells by 2D/3D interface
  engineering. \emph{Nature Communications} \textbf{2017}, \emph{8},
  15684\relax
\mciteBstWouldAddEndPuncttrue
\mciteSetBstMidEndSepPunct{\mcitedefaultmidpunct}
{\mcitedefaultendpunct}{\mcitedefaultseppunct}\relax
\EndOfBibitem
\bibitem[Gozukara~Karabag \latin{et~al.}()Gozukara~Karabag, Karabag, Gunes,
  Gao, Syzgenteva, Syzgenteva, Varlioglu~Yaylali, Shibayama, Kanda, Rafieh,
  Turnell-Ritson, Dyson, Yerci, Nazeeruddin, and Gunbas]{Karabag2023}
Gozukara~Karabag,~Z.; Karabag,~A.; Gunes,~U.; Gao,~X.-X.; Syzgenteva,~O.~A.;
  Syzgenteva,~M.~A.; Varlioglu~Yaylali,~F.; Shibayama,~N.; Kanda,~H.;
  Rafieh,~A.~I.; Turnell-Ritson,~R.~C.; Dyson,~P.~J.; Yerci,~S.;
  Nazeeruddin,~M.~K.; Gunbas,~G. Tuning 2D Perovskite Passivation: Impact of
  Electronic and Steric Effects on the Performance of 3D/2D Perovskite Solar
  Cells. \emph{Advanced Energy Materials} \emph{n/a}, 2302038\relax
\mciteBstWouldAddEndPuncttrue
\mciteSetBstMidEndSepPunct{\mcitedefaultmidpunct}
{\mcitedefaultendpunct}{\mcitedefaultseppunct}\relax
\EndOfBibitem
\bibitem[Shirzadi \latin{et~al.}(2023)Shirzadi, Ansari, Jinno, Tian, Ouellette,
  Eickemeyer, Carlsen, Van~Muyden, Kanda, Shibayama, Tirani, Gr{\"a}tzel,
  Hagfeldt, Nazeeruddin, and Dyson]{Shirzadi2023}
Shirzadi,~E.; Ansari,~F.; Jinno,~H.; Tian,~S.; Ouellette,~O.;
  Eickemeyer,~F.~T.; Carlsen,~B.; Van~Muyden,~A.; Kanda,~H.; Shibayama,~N.;
  Tirani,~F.~F.; Gr{\"a}tzel,~M.; Hagfeldt,~A.; Nazeeruddin,~M.~K.;
  Dyson,~P.~J. High-Work-Function 2D Perovskites as Passivation Agents in
  Perovskite Solar Cells. \emph{ACS Energy Letters} \textbf{2023}, \emph{8},
  3955--3961\relax
\mciteBstWouldAddEndPuncttrue
\mciteSetBstMidEndSepPunct{\mcitedefaultmidpunct}
{\mcitedefaultendpunct}{\mcitedefaultseppunct}\relax
\EndOfBibitem
\bibitem[Lin \latin{et~al.}(2018)Lin, Bai, Fang, Chen, Yang, Zheng, Tang, Liu,
  Zhao, and Huang]{Lin2018}
Lin,~Y.; Bai,~Y.; Fang,~Y.; Chen,~Z.; Yang,~S.; Zheng,~X.; Tang,~S.; Liu,~Y.;
  Zhao,~J.; Huang,~J. Enhanced Thermal Stability in Perovskite Solar Cells by
  Assembling 2D/3D Stacking Structures. \emph{The Journal of Physical Chemistry
  Letters} \textbf{2018}, \emph{9}, 654--658\relax
\mciteBstWouldAddEndPuncttrue
\mciteSetBstMidEndSepPunct{\mcitedefaultmidpunct}
{\mcitedefaultendpunct}{\mcitedefaultseppunct}\relax
\EndOfBibitem
\bibitem[Tan \latin{et~al.}(2017)Tan, Jain, Voznyy, Lan, de~Arquer, Fan,
  Quintero-Bermudez, Yuan, Zhang, Zhao, Fan, Li, Quan, Zhao, Lu, Yang,
  Hoogland, and Sargent]{Tan2017}
Tan,~H. \latin{et~al.}  Efficient and stable solution-processed planar
  perovskite solar cells via contact passivation. \emph{Science} \textbf{2017},
  \emph{355}, 722--726\relax
\mciteBstWouldAddEndPuncttrue
\mciteSetBstMidEndSepPunct{\mcitedefaultmidpunct}
{\mcitedefaultendpunct}{\mcitedefaultseppunct}\relax
\EndOfBibitem
\bibitem[Cho \latin{et~al.}(2018)Cho, Soufiani, Yun, Kim, Lee, Seidel, Deng,
  Green, Huang, and Ho-Baillie]{yongyoon2018}
Cho,~Y.; Soufiani,~A.~M.; Yun,~J.~S.; Kim,~J.; Lee,~D.~S.; Seidel,~J.;
  Deng,~X.; Green,~M.~A.; Huang,~S.; Ho-Baillie,~A. W.~Y. Mixed 3D–2D
  Passivation Treatment for Mixed-Cation Lead Mixed-Halide Perovskite Solar
  Cells for Higher Efficiency and Better Stability. \emph{Advanced Energy
  Materials} \textbf{2018}, \emph{8}, 1703392\relax
\mciteBstWouldAddEndPuncttrue
\mciteSetBstMidEndSepPunct{\mcitedefaultmidpunct}
{\mcitedefaultendpunct}{\mcitedefaultseppunct}\relax
\EndOfBibitem
\bibitem[Zhang \latin{et~al.}(2018)Zhang, Bai, Jin, Bian, Wang, Sun, Wang, and
  Liu]{Zhang2018}
Zhang,~J.; Bai,~D.; Jin,~Z.; Bian,~H.; Wang,~K.; Sun,~J.; Wang,~Q.; Liu,~S.~F.
  3D–2D–0D Interface Profiling for Record Efficiency All-Inorganic CsPbBrI2
  Perovskite Solar Cells with Superior Stability. \emph{Advanced Energy
  Materials} \textbf{2018}, \emph{8}, 1703246\relax
\mciteBstWouldAddEndPuncttrue
\mciteSetBstMidEndSepPunct{\mcitedefaultmidpunct}
{\mcitedefaultendpunct}{\mcitedefaultseppunct}\relax
\EndOfBibitem
\bibitem[Koh \latin{et~al.}(2018)Koh, Shanmugam, Guo, Lim, Filonik, Herzig,
  Müller-Buschbaum, Swamy, Chien, Mhaisalkar, and Mathews]{Koh2018}
Koh,~T.~M.; Shanmugam,~V.; Guo,~X.; Lim,~S.~S.; Filonik,~O.; Herzig,~E.~M.;
  Müller-Buschbaum,~P.; Swamy,~V.; Chien,~S.~T.; Mhaisalkar,~S.~G.;
  Mathews,~N. Enhancing moisture tolerance in efficient hybrid 3D/2D perovskite
  photovoltaics. \emph{J. Mater. Chem. A} \textbf{2018}, \emph{6},
  2122--2128\relax
\mciteBstWouldAddEndPuncttrue
\mciteSetBstMidEndSepPunct{\mcitedefaultmidpunct}
{\mcitedefaultendpunct}{\mcitedefaultseppunct}\relax
\EndOfBibitem
\bibitem[Blancon \latin{et~al.}(2017)Blancon, Tsai, Nie, Stoumpos, Pedesseau,
  Katan, Kepenekian, Soe, Appavoo, Sfeir, Tretiak, Ajayan, Kanatzidis, Even,
  Crochet, and Mohite]{Blancon2017}
Blancon,~J.-C. \latin{et~al.}  Extremely efficient internal exciton
  dissociation through edge states in layered 2D perovskites. \emph{Science}
  \textbf{2017}, \emph{355}, 1288--1292\relax
\mciteBstWouldAddEndPuncttrue
\mciteSetBstMidEndSepPunct{\mcitedefaultmidpunct}
{\mcitedefaultendpunct}{\mcitedefaultseppunct}\relax
\EndOfBibitem
\bibitem[Yu and Zunger(2012)Yu, and Zunger]{Yu2012}
Yu,~L.; Zunger,~A. Identification of Potential Photovoltaic Absorbers Based on
  First-Principles Spectroscopic Screening of Materials. \emph{Phys. Rev.
  Lett.} \textbf{2012}, \emph{108}, 068701\relax
\mciteBstWouldAddEndPuncttrue
\mciteSetBstMidEndSepPunct{\mcitedefaultmidpunct}
{\mcitedefaultendpunct}{\mcitedefaultseppunct}\relax
\EndOfBibitem
\end{mcitethebibliography}
\end{document}